# An Overview of 5G Advanced Evolution in 3GPP Release 18

Xingqin Lin

Ericsson

Email: xingqin.lin@ericsson.com

*Abstract*— **The 3rd generation partnership project (3GPP) radio access network (RAN) plenary recently approved a work package for its Release 18, representing a major evolution and branded as the first release of 5G Advanced. The work package includes diverse study or work items that will significantly boost 5G performance and address a wide variety of new use cases. In particular, 3GPP Release 18 will embrace artificial intelligence and machine learning technologies to provide data-driven, intelligent network solutions. This article provides an overview of the 5G Advanced evolution in 3GPP Release 18, which is anticipated to trigger a paradigm shift and have a profound impact on future wireless networks.**

## I. INTRODUCTION

The 3rd generation partnership project (3GPP) completed the first release of the fifth-generation (5G) of mobile communications in its Release-15 in June 2018, which laid the basis for commercial 5G deployments worldwide [1]. Since then, 3GPP has been working on evolving 5G technology in its Releases 16 and 17 to improve performance further and address new use cases [2]. 3GPP recently approved the work package for its Release 18, which will mark the start of 5G Advanced evolution. In this article, we provide an overview of the 5G Advanced evolution in 3GPP Release 18.

Figure 1 shows 3GPP's 5G evolution roadmap from 5G to 5G Advanced. In Release 15, 3GPP specified a new 5G air interface, known as New Radio (NR), aiming to address a variety of usage scenarios from enhanced mobile broadband (eMBB) to ultra-reliable low-latency communications (URLLC) to massive machine type communications (mMTC). NR supports both non-standalone (NSA) operation, which utilizes Long Term Evolution (LTE) for initial access and mobility handling, and standalone (SA) operation without relying on LTE. Key NR features include high-frequency operation and spectrum flexibility, ultra-lean design, forward compatibility, flexible duplex schemes, low-latency support, advanced antenna technologies, among others [3].

Release 16, the first step in the 5G evolution, introduces several major enhancements that improve existing features and address new use cases and deployment scenarios. Key improvements of existing features include multiple-input multiple-output (MIMO) and beamforming enhancements, enhanced support of dynamic spectrum sharing (DSS), reduced latency in dual connectivity (DC) and carrier aggregation (CA), and user equipment (UE) power saving. The main new use cases and deployment scenarios addressed in Release 16 include enhanced support of industrial Internet of Things (IIoT) and URLLC, operation in unlicensed spectrum, vehicle-to-anything (V2X) communications, positioning services, and integrated access and backhaul (IAB). 3GPP's submission to International Mobile Telecommunications-2020 (IMT-2020) contained Release 15 and Release 16 functionality. International Telecommunication Union Radiocommunication Sector (ITU-R) has approved the 3GPP's submission as 5G technology in 2020.

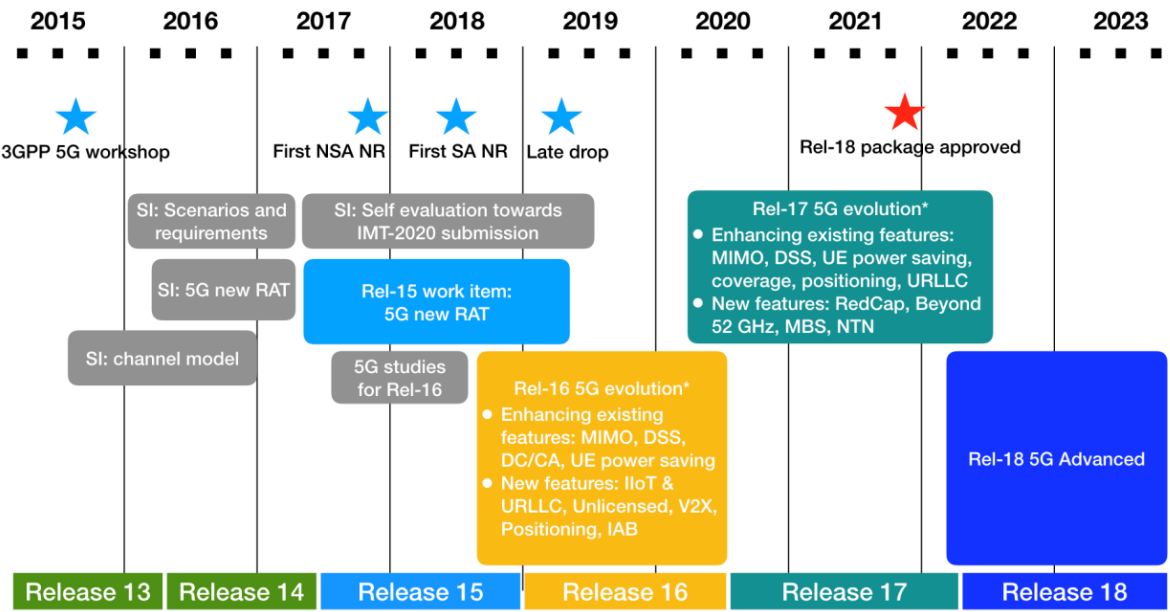

**Figure 1: 3GPP's 5G evolution roadmap from 5G to 5G Advanced (indicative).**

3GPP continues 5G evolution in its Release 17. Existing features, such as MIMO, DSS, UE power saving, coverage, positioning, and URLLC, are further enhanced. The main new use cases and deployment scenarios addressed in Release 17 include support for reduced capability (RedCap) UE, operation in frequency bands beyond 52 GHz, multicast and broadcast service (MBS), and non-terrestrial networks (NTNs). It is worth noticing that Release 17 is the first time that 3GPP completes a full release of work all by electronic meetings.

Release 18 will be the start of work on 5G Advanced. The discussion on the scope of Release 18 can be traced back to the 3GPP radio access network (RAN) Release-18 workshop, which attracted more than 500 proposals. After a 6-month intense discussion, 3GPP approved its Release-18 package at the December 2021 RAN plenary meeting [4]. The package includes 27 diverse study or work items that will further boost network performance and address new use cases. In particular, the package features work on embracing artificial intelligence (AI) and machine learning (ML) technologies in the evolution of 5G Advanced.

This article aims to provide an overview of the 5G Advanced evolution in 3GPP Release 18 at a level accessible to a general audience working in the wireless communications and networking communities. To make the scope of the 3GPP Release 18 package more easily digestible, we categorize the approved 27 items into five groups in this article and present them in Sections II, III, IV, V, and VI, respectively. Note that 3GPP typically specifies features in a manner agnostic to specific services or use cases. The categorization in this article does not mean that a feature discussed in a particular group cannot be applied to the other groups; on the contrary, it is up to implementation and deployment to choose and combine the specified features.

## II. FURTHER ENHANCED 5G PERFORMANCE

3GPP will continue to study and add functionality in Release 18 to enhance 5G performance by working in the areas of network energy savings, coverage, mobility support, MIMO evolution, MBS, and positioning, as illustrated in Figure 2.

The impetus for saving network energy has grown remarkably as energy cost has become a significant part of network operating expenses (OPEX) [5]. Besides, reductions in network energy consumption are imperative for operators to ultimately achieve the sustainability objective of net zero to help combat climate change. 5G NR is significantly more energy efficient in terms of energy consumption per bit than the previous generations of mobile technologies. However, due to the increasingly dense deployment of 5G networks, massive MIMO, much larger bandwidths, and support of more frequency bands, 5G networks would lead to higher energy consumption if proper energy-saving measures are not applied. Therefore, 3GPP approves a study on network energy savings for NR in Release 18. The study aims to define a network energy consumption model for a base station (BS), develop an evaluation methodology, and identify key performance indicators (KPIs). The study will then investigate techniques to improve network energy savings in targeted deployment scenarios.

Coverage is another key consideration for operators as it directly impacts service quality, capital expenditures (CAPEX), and OPEX in commercial deployments. Uplink coverage is often found to be the bottleneck in commercial deployments. In Release 17, 3GPP has added techniques to extend uplink coverage for the physical uplink shared channel (PUSCH) and physical uplink control channel (PUCCH). In Release 18, 3GPP plans to enhance the coverage of physical random access channel (PRACH) and study techniques to increase UE power limit for CA and DC and reduce maximum power reduction or peak-average-power ratio. In addition, dynamic switching between cyclic-prefix orthogonal frequency-division multiplexing (CP-OFDM) and discrete Fourier transform (DFT) spread OFDM (DFT-S-OFDM) in the uplink will be studied.

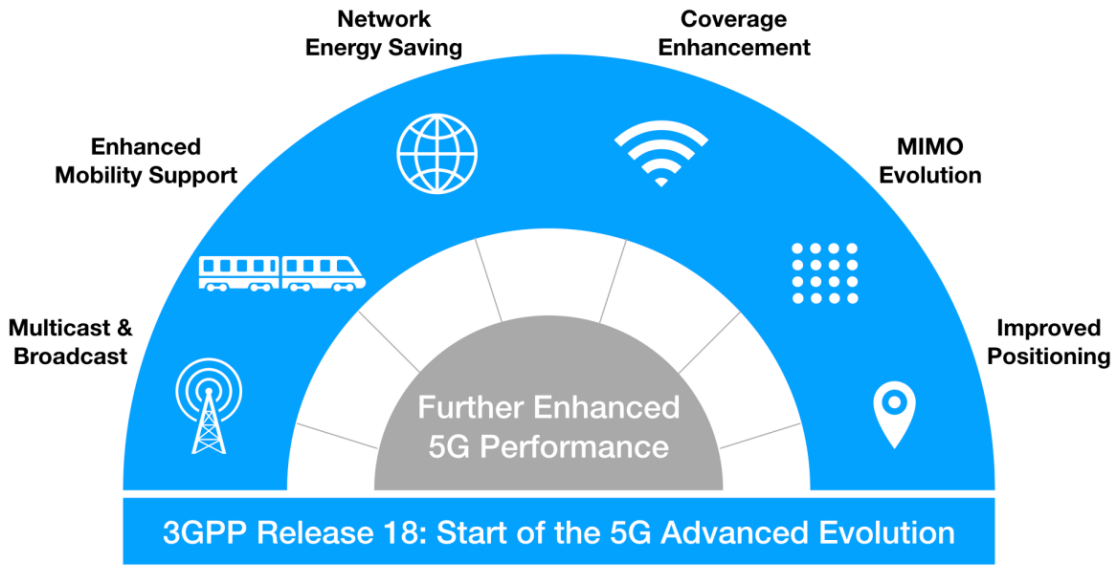

**Figure 2: An illustration of the key 3GPP Release-18 features that will further enhance 5G performance.**

Mobility support is a distinct feature of mobile networks which offers service continuity to moving UEs [6]. When a UE moves from the coverage area of one cell to another, a handover process is initiated to change the serving cell for the UE. The serving cell change is currently based on layer 3 (L3) measurements and radio resource control (RRC) signaling. To reduce the latency, overhead, and interruption time associated with L3-based mobility management, 3GPP Release 18 will introduce mechanisms and procedures for layer 1 (L1)/layer 2 (L2) based inter-cell mobility. Another major area for Release-18 mobility work is to enhance further the support of conditional handover. In conditional handover, UE receives a handover command with a condition from the network and does not apply the command until the condition is satisfied.

MIMO evolution continues in 3GPP Release 18 [7]. It has been observed in commercial deployments that UE with medium or high mobility experiences significant performance loss in multi-user MIMO (MU-MIMO) scenarios, partially due to outdated channel state information (CSI). Potential CSI reporting enhancements will be explored to improve performance for UE with medium or high mobility. Also, the work will study a larger number of orthogonal demodulation reference signal (DMRS) ports for MU-MIMO. A unified TCI framework has been introduced in Release 17 for single transmission-reception point (TRP). 3GPP will extend the TCI framework to multi-TRP scenarios and study coherent joint transmission, two timing advances, and enhanced uplink power control for multi-TRP. To better support advanced UEs such as customer premise equipment (CPE), fixed wireless access (FWA) devices, and vehicular UEs, support of eight antenna ports in uplink and simultaneous multi-panel uplink transmission will be investigated.

MBS is a feature dedicated to delivering multicast and broadcast services efficiently. Use case examples include TV broadcasting, live video, software updates, and public safety usages. 3GPP specified the basic support for MBS in Release 17, enabling multicast transmission to UEs in RRC connected state and broadcast transmission to UEs in all RRC states. 3GPP Release 18 will extend the multicast support to UEs in RRC inactive state, introduce enhancements to enable UEs in RRC connected state to receive broadcast service and unicast service simultaneously, and study mechanisms to improve resource efficiency in RAN sharing scenarios.

Positioning is a valuable service that can find applications in diverse 5G use cases. While Release-15 NR supported positioning, e.g., by using LTE positioning in NSA operation, Release-16 NR much enhanced the positioning support with a range of positioning methods, including both downlink-based and uplink-based positioning. Release-17 NR introduced additional enhancements to reduce latency for time-critical use cases such as remote control, deliver positioning accuracy down to the level of 20-30 cm for use cases such as factory automation, and improve integrity protection of the location information. 3GPP Release 18 will investigate solutions to further improve accuracy, integrity, and power efficiency in positioning, study sidelink positioning, and investigate positioning support for RedCap devices.

## III. FLEXIBLE SPECTRUM USE

Spectrum is a scarce resource that needs to be used efficiently to achieve the highest societal benefit. 3GPP Release 18 will introduce further enhancements to enable more flexible and efficient spectrum use for 5G deployments in various scenarios with different spectrum allocations. The main enabling features are illustrated in Figure 3 and described below.

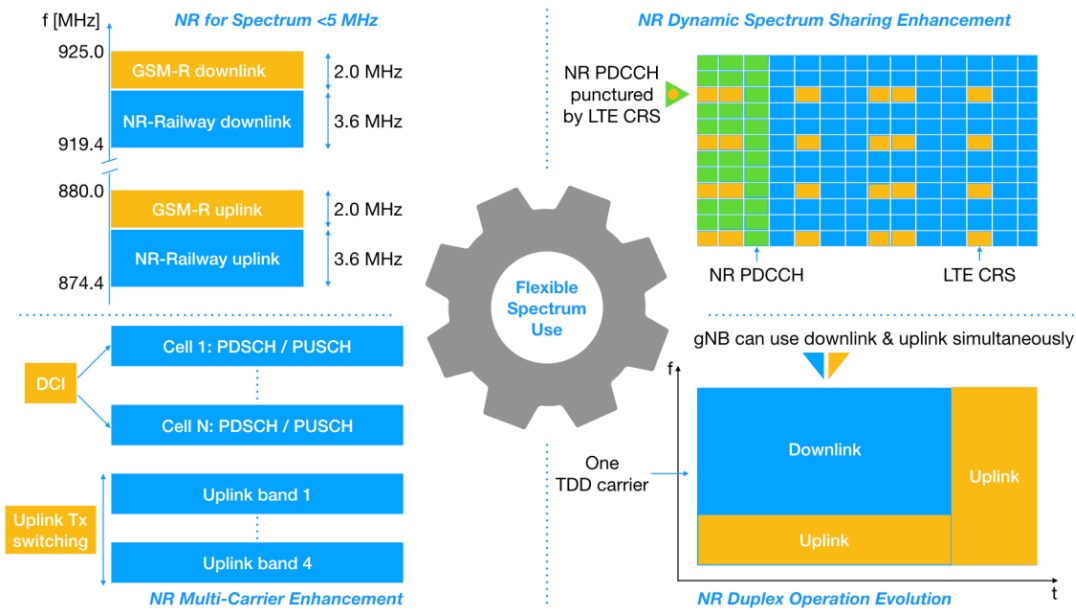

**Figure 3: An illustration of the main 3GPP Release-18 features that will enable more flexible spectrum use in 5G deployments.**

The minimum channel bandwidth supported by the current 5G NR specifications is 5 MHz. There are, however, growing interests in deploying NR in a dedicated spectrum with less than 5 MHz bandwidth available for NR. For example, railway communication in Europe, currently based on the global system for mobile communications-railway (GSM-R), will be migrated to the future railway mobile communication system (FRMCS) by using the harmonized 900 MHz spectrum block (2 x 5.6 MHz frequency division duplex (FDD)) [8]. However, FRMCS will need to co-exist with the legacy GSM-R for about ten years from 2025 onwards. Keeping the GSM-R system operational will require a portion of the spectrum block reserved for the legacy system, making the spectrum available to NR deployment for FRMCS less than 5 MHz. In Release 18, 3GPP will introduce necessary changes to NR to support deployments in spectrum allocations less than 5 MHz.

DSS is a crucial 5G feature that enables a BS to use a shared spectrum to provide connectivity to both LTE UE and NR UE. This facilitates the migration of the spectrum from LTE to NR. In DSS, the physical downlink control channel (PDCCH) capacity is a bottleneck, because NR PDCCH and LTE PDCCH need to share the first three OFDM symbols within a slot and NR PDCCH cannot use symbols overlapping with LTE cell-specific reference signal (CRS). In Release 18, 3GPP will study the possibility of allowing NR PDCCH to be transmitted in symbols overlapping with LTE CRS to increase the PDCCH capacity for DSS. Also, the feature of configuring UE with multiple LTE CRS rate matching patterns in multi-TRP will be made available to single TRP scenario to help mitigate inter-cell interference.

In general, it is expected that more and more spectrum bands used by the previous generations of mobile networks will be re-farmed to be used for 5G Advanced. The available spectrum blocks will likely be fragmented and scattered in different frequencies. Aiming to efficiently utilize all the available spectrum blocks flexibly, 3GPP Release 18 will introduce a functionality that allows a single downlink control information (DCI) to schedule multiple physical downlink shared channels (PDSCHs) or PUSCHs across carriers and study enhancements for multi-carrier uplink operation.

Time division duplex (TDD) is widely used in commercial 5G deployments. Downlink and uplink can use the entire frequency spectrum in TDD, but the time domain resource is divided between downlink and uplink. NR supports TDD with semi-statically configured uplink/downlink configuration and dynamic TDD. If the time domain resource designated to uplink is limited, network performance such as coverage, capacity, and latency may be impacted. To enable more flexible spectrum use, 3GPP Release 18 will study the feasibility of allowing the co-existence of downlink and uplink at the same time within a conventional TDD band. In other words, the duplex will become a mix of TDD and FDD. From a practical standpoint, since full duplex is not mature enough for commercial deployments, the study will restrict its scope to subband non-overlapping full duplex at the 5G Node B (gNB) side. In other words, a conventional TDD band is divided into non-overlapping subbands: some are designated for uplink while others are designated for downlink. But half duplex is kept at the UE side. Besides, 3GPP Release 18 will study cross link interference (CLI) handling to better support dynamic TDD in commercial deployments.

## IV. DIVERSE 5G DEVICES

5G needs to serve diverse devices across eMBB, URLLC, and mMTC usage scenarios. 3GPP Release 18 will continue to study and introduce tailored functionalities to enhance and expand 5G capability to serve not only smartphones but also other diverse 5G devices, such as extended reality (XR) and cloud gaming devices, low-complexity UEs, vehicular devices, and uncrewed aerial vehicles (UAVs), as illustrated in Figure 4.

### A. 5G Consumer Devices

A smartphone typically supports both 3GPP radio access technologies (RATs), e.g., LTE and NR, and non-3GPP RATs, e.g., WiFi. The coexistence of a 3GPP RAT and non-3GPP RATs may lead to interference and cause an internal issue in the device. In-device coexistence (IDC) is a feature that allows UE to indicate to its serving gNB that there is such a coexistence issue in the UE. The gNB can utilize the indicated information to restrict radio resource usage to help resolve the UE's internal issue. The current IDC solution in NR has limited functionality. 3GPP Release 18 will enhance the feature by allowing UE to indicate the affected frequencies with finer granularity and its preferred TDD pattern.

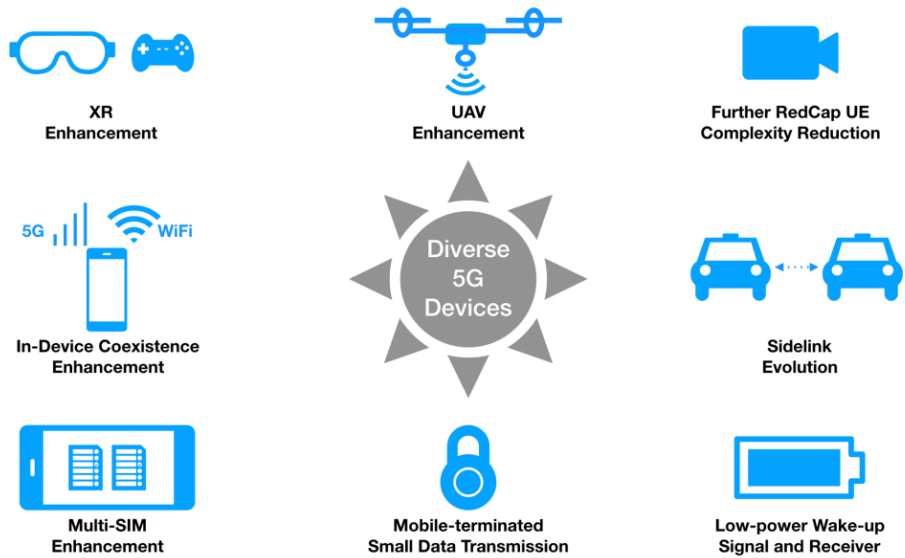

**Figure 4: An illustration of the main 3GPP Release-18 features that will expand 5G capability to serve diverse devices.**

A multi-subscriber identity module (SIM) phone can be equipped with two independent SIM cards. When the UE is only connected to one network (e.g., network A) using one SIM card, the UE's hardware resources can be dedicated to the communication with network A. Once the UE starts to connect to another network (e.g., network B) using the other SIM card, some of the hardware resources need to be used for communication with network B, but this is not known to network A. 3GPP Release 18 will enable the multi-SIM UE to indicate its preferred capability restriction to network A when the UE needs to communicate with network B.

5G connected XR and cloud gaming devices are anticipated to proliferate in years to come. In Release 17, 3GPP has conducted an evaluation study on XR to assess their performance when connected by 5G. In Release 18, 3GPP will study enhancements to better support XR and cloud gaming devices in NR networks. The traffic generated in XR and cloud gaming use cases is often quasi-periodic and requires high data rates and bounded latency simultaneously. To efficiently serve such type of traffic, 3GPP will investigate resource allocation and scheduling mechanisms that can improve capacity for XR and cloud gaming devices. Since many XR and cloud gaming devices have limited battery power, Release 18 will also study UE power-saving techniques to accommodate XR and cloud gaming service characteristics. Besides, 3GPP will investigate how to make RAN more XR-aware, including identifying what application information is beneficial for RAN to be aware of and how to utilize the information in RAN to handle XR and cloud gaming traffic.

### B. 5G Devices for Verticals

While serving high-end eMBB devices is vital in 5G NR networks, there have been growing interests in providing 5G connectivity to RedCap devices [9]. Use cases include industrial sensors, video surveillance, and wearables, which require low UE complexity and low UE power consumption. 3GPP Release 17 has established a basic framework for supporting RedCap devices in NR networks. Built on the Release-17 foundation, 3GPP Release-18 will study enhancements to support RedCap devices of even lower complexity, targeting UE bandwidth reduction to 5 MHz and peak data rate reduction to 10 Mbps in the frequency range 1 (FR1).

Devices such as industrial sensors and wearables have small form-factor and are power sensitive. UE power consumption depends on the configuration of discontinuous reception (DRX) cycle, e.g., paging cycle for UE in idle mode, because UE needs to wake up once per DRX cycle. When there is no signaling or data traffic, UE would wake up in vain and waste power. Therefore, it is beneficial to wake up UE when it is triggered. Specifically, UE can be equipped with a separate receiver with ultra-low power consumption, which can monitor a wake-up signal. Upon detecting the wake-up signal, UE can trigger its main communication radio from OFF state or deep sleep mode to become active. 3GPP Release 18 will study low-power wake-up receiver architectures, the design of a wake-up signal, and procedure and protocol changes to support the wake-up receiver and signal.

In many scenarios, low-power devices only need to support infrequent, small data transmissions. Transmitting such small data in RRC connected mode would lead to signaling overhead and delay. To reduce signaling overhead and latency, 3GPP Release 17 introduced support for mobile originated small data transmission in RRC inactive mode. In Release 18, 3GPP will introduce a similar feature to support mobile terminated small data transmission in RRC inactive mode.

3GPP continues to add functionality in Release 18 to enhance sidelink communication to support vehicular devices used in V2X services [10]. One evolution objective is to increase sidelink data rate by adding the CA feature in sidelink communication, extending sidelink operation to unlicensed spectrum, and enhancing sidelink support in the frequency range 2 (FR2). Besides, 3GPP will study mechanisms to support LTE V2X and NR V2X devices co-existing in the same frequency channel.

In addition to enhancing support for V2X devices on the ground, 3GPP Release 18 will introduce 5G NR support for devices onboard aerial vehicles, i.e., UAVs, whose applications are proliferating across many industries. 3GPP has introduced 4G LTE support for UAVs back in Release 15, including enhanced measurement reports, reporting of UAV height, location and speed, flight path reporting, and signaling support for subscription-based aerial UE identification [11]. These features will be introduced into 5G NR specifications as appropriate. Besides, the work will study enhancements for broadcasting UAV identification and signaling to indicate UAV beamforming capabilities.

## V. EVOLVED NETWORK TOPOLOGY

Coverage is crucial for operators to provide competitive mobile services to their customers. Deploying full-stack BSs in coverage holes is a common way to improve network coverage. Still, it may not always be possible due to, e.g., lack of backhaul, economically not viable, among others. This calls for flexible network topology by increasing the resiliency of the split next-generation RAN (NG-RAN) architecture, exploiting diverse nodes such as IAB nodes, radio frequency (RF) repeaters, and relays, and integration with NTNs. The key enabling features are illustrated in Figure 5 and described below.

NG-RAN – the RAN for 5G – supports splitting a gNB into two parts: a centralized unit (CU) and one or more distributed units (DUs) [12]. The gNB-CU can be further split into two parts: a control plane (CP) part and one or more user plane (UP) parts. The logical gNB-CU-CP hosts the RRC and the CP part of the packet data convergence protocol (PDCP) of the gNB-CU. Failures at the gNB-CU-CP may disrupt its connected gNB-CU-UPs and gNB-DUs, which in turn can impact the served UEs. To enhance the resiliency of gNB-CU-CP, 3GPP

Release 18 will study and identify gNB-CU-CP failure scenarios.

IAB is a wireless backhaul solution based on NR. It serves as an alternative to fiber backhaul and thus can facilitate the deployment of mobile networks. An IAB node may consist of two parts: mobile-termination (MT) and DU. The MT connects the IAB node to its donor node, while the DU part can serve UEs and connect to other IAB nodes to create multi-hop wireless backhauling. 3GPP Release 18 continues to enhance NR-based IAB, focusing on the scenario where mobile IAB nodes are mounted on vehicles and provide 5G connectivity to UEs. To support mobile IAB nodes, the work will specify topology adaptation procedures, mobility enhancements for IAB nodes and the associated UEs, and interference mitigation.

RF repeaters can help extend network coverage and have been used in commercial deployments. However, they have limited functionality as they simply amplify and forward the signals they receive. 3GPP Release 18 will study NR network-controlled repeater, which not only can amplify and forward the received signals but also can receive side control information from the gNB. The examples of side control information include beamforming information, TDD configuration, power control information, among others. The study will identify which side control information is necessary, investigate needed L1/L2 signaling, and look into how to identify and authorize such network-controlled repeaters.

Network coverage can also be extended by using a UE as a sidelink relay. The sidelink relay can be either a UE-to-network relay, where the relay UE connects a remote UE to the network, or a UE-to-UE relay, where the relay UE connects a first remote UE to a second remote UE. In Release 17, 3GPP specified basic functionality support for UE-to-network relay. 3GPP Release 18 will enhance the support of UE-to-network relay, add support for UE-to-UE relay, and study multi-path support where a UE is connected to a gNB using one direct path and one indirect path via a sidelink relay.

NTNs utilize satellites or high-altitude platforms (e.g., airplanes, balloons, and airships) to offer connectivity services [13]. They can complement terrestrial networks by providing coverage in remote areas where terrestrial coverage is unavailable. In Release 17, 3GPP has introduced a set of basic features to enable NR operation over NTNs in FR1. 3GPP Release 18 will enhance NR operation over NTNs by improving coverage for handheld terminals, studying deployment above 10 GHz, investigating regulatory requirements for network-verified UE location, and addressing mobility and service continuity between a terrestrial network and an NTN as well as between different NTNs.

## VI. DATA DRIVEN AND AI-POWERED 5G

5G networks are becoming increasingly complex while generating humongous data. It is crucial to be data-driven and leverage AI techniques to manage the 5G networks efficiently. 3GPP Release 18 will not only enhance existing data collection features but also examine how AI techniques can improve air interface functions [14].

In Release 17, 3GPP conducted a study on AI-enabled NG-RAN. The study investigated high-level principles, functional framework, potential use cases, and associated solutions for AI-enabled RAN intelligence [15]. 3GPP Release 18 will specify data collection enhancements and signaling support for a set of selective AI-based use cases, including network energy saving, load balancing, and mobility optimization.

3GPP Release 18 will also enhance NR data collection within the scope of the self-organizing network (SON)/minimization of drive testing (MDT). SON automates RAN planning, configuration, management, optimization, and healing, thereby minimizing human intervention. MDT enables operators to configure normal UEs to collect and report measurement data to reduce traditional drive tests. The Release-18 work will address SON features leftover from Release 17 and data collection for random access channel (RACH) optimization.

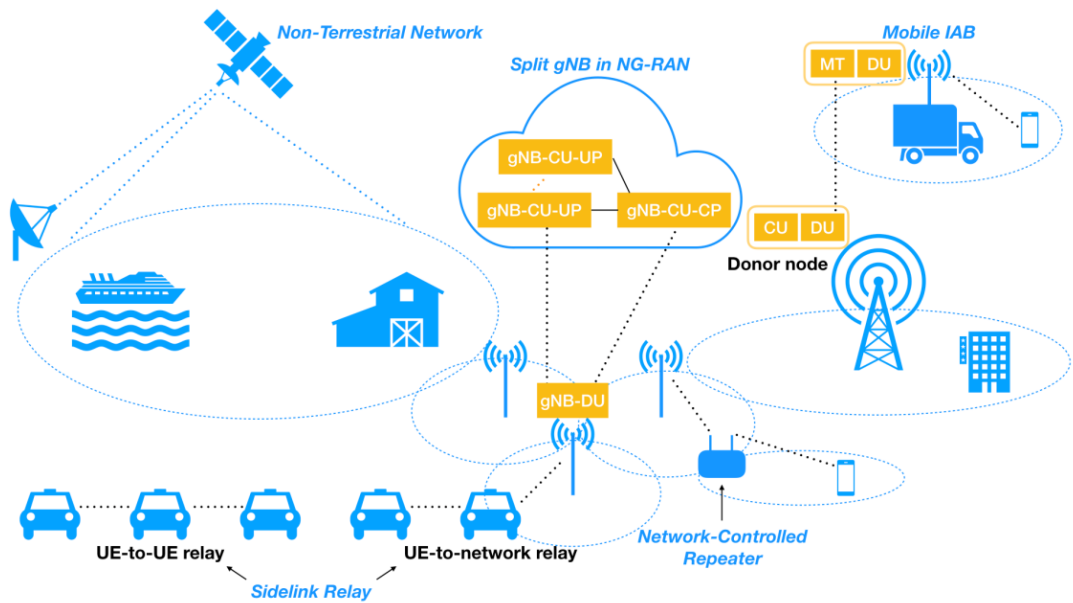

**Figure 5: An illustration of the key 3GPP Release-18 features that will enable a more flexible 5G network topology.**

| Category | Approved item | Study/work item | Responsible groups |
|---|---|---|---|
| **Further enhanced 5G performance** | RP-213554: Study on network energy savings for NR | Study item | RAN 1, 2, 3 |
| | RP-213579: Further NR coverage enhancements | Work item | RAN 1, 2, 4 |
| | RP-213565: Further NR mobility enhancements | Work item | RAN 2, 1, 3, 4 |
| | RP-213598: NR MIMO evolution for downlink and uplink | Work item | RAN 1, 2, 4 |
| | RP-213568: Enhancements of NR multicast and broadcast services | Work item | RAN 2, 3 |
| | RP-213588: Study on expanded and improved NR positioning | Study item | RAN 1, 2, 3, 4 |
| **Flexible spectrum use** | RP-213603: NR support for dedicated spectrum less than 5MHz for FR1 | Work item | RAN 4, 1, 2 |
| | RP-213575: Enhancement of NR dynamic spectrum sharing (DSS) | Work item | RAN 1, 2, 4 |
| | RP-213577: Multi-carrier enhancements for NR | Work item | RAN 1, 2, 4 |
| | RP-213591: Study on evolution of NR duplex operation | Study item | RAN 1, 4 |
| **Diverse 5G devices** | RP-213589: In-device co-existence (IDC) enhancements for NR and MR-DC | Work item | RAN 2, 4 |
| | RP-213584: Dual Tx/Rx multi-SIM for NR | Work item | RAN 2 |
| | RP-213587: Study on XR enhancements for NR | Study item | RAN 2, 1 |
| | RP-213661: Study on further NR RedCap UE complexity reduction | Study item | RAN 1 |
| | RP-213645: Study on low-power wake-up signal and receiver for NR | Study item | RAN 1, 2, 4 |
| | RP-213583: Mobile terminated-small data transmission for NR | Work item | RAN 2, 3 |
| | RP-213678: NR sidelink evolution | Work item | RAN 1, 2, 4 |
| | RP-213600: NR support for UAV (uncrewed aerial vehicles) | Work item | RAN 2, 1 |
| **Evolved network topology** | RP-213677: Study on enhancement for resiliency of gNB-CU-CP | Study item | RAN 3 |
| | RP-213601: Mobile IAB | Work item | RAN 3, 2, 4 |
| | RP-213700: Study on NR network-controlled repeaters | Study item | RAN 1, 2, 3 |
| | RP-213585: NR sidelink relay enhancements | Work item | RAN 2, 3, 4 |
| | RP-213690: NR NTN (non-terrestrial networks) enhancements | Work item | RAN 2, 1, 3, 4 |
| **Data-driven and AI-powered 5G** | RP-213602: Artificial intelligence (AI)/machine learning (ML) for NG-RAN | Work item | RAN 3 |
| | RP-213553: Further enhancement of data collection for SON/MDT in NR and EN-DC | Work item | RAN 3, 2 |
| | RP-213594: Enhancement on NR QoE management and optimization for diverse services | Work item | RAN 3, 2 |
| | RP-213599: Study on artificial intelligence (AI)/machine learning (ML) for NR air interface | Study item | RAN 1, 2, 4 |

**Table 1: A summary of 3GPP RAN Release-18 package approved at the December 2021 RAN plenary meeting.**

Besides data collection of radio-related measurement and information, the 5G NR quality of experience (QoE) management framework enables measurement data collection at the application layer. QoE reflects the user's satisfaction level with a service provided by mobile networks. A QoE report contains data of QoE metrics collected at the application layer. Examples of QoE metrics in multimedia telephony service for internet protocol multimedia subsystem (IMS) include frame rate, jitter duration, round-trip time, average codec bit rate, among others. In Release 17, 3GPP specified basic mechanisms for NR QoE management. 3GPP Release 18 will enhance the NR QoE framework to support new types of 5G services, such as XR, cloud gaming, and MBS.

3GPP Release 18 will study AI/ML for NR air interface to enhance performance or reduce complexity/overhead. The study will establish a common AI/ML framework, identify areas where AI/ML can improve air interface functions, investigate how to describe and characterize AI/ML models, evaluate AI/ML techniques to understand their gains and complexity, and assess standardization impact. To achieve these objectives, 3GPP will focus on a set of selective use cases, including CSI feedback, beam management, and positioning. The study is expected to pave the way for other use cases leveraging AI/ML techniques in the air interface.

## VII. CONCLUSIONS

Since 3GPP completed the first release of 5G in its Release 15, the 5G evolution has progressed swiftly in 3GPP Releases 16 and 17. 3GPP Release 18 – the first release of 5G Advanced – represents the next major evolution of 5G technologies. As overviewed in this article, 3GPP Release 18 will further enhance 5G performance, introduce features to enable more flexible and efficient spectrum use, advance the support of diverse devices, evolve network topology to facilitate different deployments, and provide data-driven, intelligent network solutions. Table 1 summarizes the 3GPP Release-18 package discussed in this article.

While we are just embarking on the evolution journey of 5G Advanced, research on the sixth-generation (6G) of mobile communications is ramping up. The 6G standardization is expected to start in 3GPP around 2025. We anticipate that the innovative works conducted in 5G Advanced, such as embracing AI and ML technologies, will trigger a paradigm shift, lay a strong foundation for 6G design, and create a profound impact on future wireless networks.